\documentstyle[12pt,psfig,epsf]{article}
\newcommand{\be}{\begin{equation}}
\newcommand{\ee}{\end{equation}}

\begin{document}
\hfill  NRCPS-HE-99-13

\hfill  NTUA-99/19

\vspace{24pt}
\begin{center}
{\large \bf Vacuum Structure of Gauge Theory on Lattice with 
Two Parallel Plaquette Action }

\vspace{24pt}
{\sl K.Farakos}\\
Physics Department, National Technical University, \\
Zografou Campus, 15780 Athens, Greece\\
{\tt email:kfarakos@central.ntua.gr}
\vspace{1cm}

{\sl G.K.Savvidy}

National Research Center Demokritos,\\
Ag. Paraskevi, GR-15310 Athens, Greece \\
{\tt email:savvidy@argo.nrcps.ariadne-t.gr}
\end{center}
\vspace{60pt}

\centerline{{\bf Abstract}}

\vspace{12pt}
\noindent
We perform Monte Carlo simulations of a lattice gauge system with 
an action which contains two parallel plaquettes. 
The action is defined as a product of gauge group 
variables over two parallel plaquettes belonging to a given three-dimensional 
cube. The peculiar property of this system is that it has strong degeneracy 
of the vacuum state inherited from corresponding gonihedric $Z_2$ gauge 
spin system. These vacuua are well separated and can 
not be connected by a gauge transformation. We measure different 
observables in these vacuua and compare their properties.


\newpage

\pagestyle{plain}
\section{Introduction}


In this article we  will report some observation that we have made in 
gauge 
invariant model defined on a lattice with an action which contains two 
parallel plaquettes \cite{sav}. 
This action is non-Abelian generalization of the $Z_2$ gauge spin system 
which was introduced as a lattice realization of random surfaces with 
gonihedric action \cite{sav1}. It is defined as a product of gauge group 
variables over two parallel plaquettes belonging to a given three-dimensional 
cube of the lattice \cite{sav}
\be
H_{gonihedric} = \frac{n}{g^{2}} \sum_{\{ parallel~plaquettes \}}
\{ 1- (1/n^2) Re~ Tr  U_{plaq} \cdot Re ~Tr U_{plaq}  \} ,      \label{linear}
\ee
where $U_{plaq}$ is a product of gauge group elements  $U_{ij}$ defined on
a link $<i,j>$ over all sites of a plaquette and $U_{ij} = exp(i g A_{\mu} a)$. 
As one can see the action contains a product of two traces taken over two 
parallel plaquettes belonging to a 3D-cube. The summation is over all pairs
of parallel plaquettes \footnote{The model with an action which is 
defined as a sum of contributions over all closed loops made up with 
six non-repeated links has been considered in \cite{carlo}. }.  

It is easy to see that for smooth classical 
fields the action reduces to the original action for non-Abelian fields 
as it takes place for the Wilson action \cite{wils,creutz}
\be
H = \frac{2n}{g^{2}} \sum_{\{plaquettes \}}
(1- (1/n) Re~Tr U_{plaq} ) .     \label{area}
\ee
Indeed the matrix product of four $U's$ around plaquette $\mu,\nu = 1,2$ 
becomes  
$$
Re~Tr U_{plaq} = Re~Tr~  exp (i g a^2 F_{12} +...)
$$
and after expanding fields around the center $x_{\mu}$ of the 3D-cube we 
will get 
$$
1 - (1/n^2)(n - \frac{g^2 a ^4 }{2} Tr F^{2}_{12}(x_{\mu} + 
\frac{a}{2}\delta_{\mu 3}))(n - \frac{g^2 a ^4 }{2} Tr F^{2}_{12}(x_{\mu} + 
\frac{a}{2} \delta_{\mu 3})) \approx \frac{g^2 a ^4}{n} Tr F^{2}_{12}
$$
therefore it reproduces the classical action for pure Yang-Mills fields. 

Thus on a classical level both models (\ref{linear}) and (\ref{area}) 
are equivalent. But on a quantum level it is 
not  obvious at all. The reason is that the vacuum structure of the model 
(\ref{linear}) is drastically different \cite{sav,weg}. In 
addition to the trivial vacuum configuration when all plaquette variables
are equal to one 
\be
\frac{1}{2}Tr~U_{plaq} = + 1, \label{plus}
\ee
there are $2^{dN}$ different vacuum configurations \cite{sav,weg} with frustrated 
plaquettes $\frac{1}{2}Tr~U_{plaq}=-1$. One of these 
vacuua can be easily described as a configuration with all 
plaquettes  frustrated 
\be
\frac{1}{2}Tr~U_{plaq} = -1. \label{minus}
\ee
It is obvious that it is impossible to connect vacuum configurations 
(\ref{plus}) and (\ref{minus}) by a gauge transformation.
The riach vacuum structure inherited from the "parent" $Z_{2}$ gauge spin 
system with the Hamiltonian \cite{sav}
\be
H_{gonihedric}= -\frac{1}{g^{2}} \sum_{\{parallel~plaquettes\}}
(\sigma\sigma\sigma\sigma)~(\sigma\sigma\sigma\sigma)  \label{parallel}
\ee
In both models (\ref{parallel}) and (\ref{linear}) we have the same vacuum 
structure: towers of frustrated plaquettes describe different vacuua
\cite{weg}. The difference between $SU(2)$ and $Z_2$ systems is that in the 
model (\ref{linear}) we have a continuous Lie gauge group and therefore 
continuous fluctuations of gauge fields around these vacuua. 

The problem of our main concern is 
the physics in these vacuua and its relevance to the continuum limit. It is also
interesting to understand: what is the remnant of this riach vacuum structure in 
the continuum limit. Is there any dilaton-like field which describes different 
vacuua in the continuum limit or not? And if there is a continuum limit in all
these vacuua, then what is the difference between them? 
In the next section we will present results of simple analytical 
consideration of 
the model and the results of the Monte Carlo simulations for the $SU(2)$ gauge 
group which are  performed on three-dimensional lattices for an  obvious 
simplification \cite{kostas,teper}.

\section{High and low temperature expansion of average action}

As usually, all observables can be constructed as a product of gauge variables 
over closed loops and we shall consider first the average  action.
For the Wilson action (\ref{area}) a simple one-plaquette average is equal to 
\cite{creutz}
$$
P ~=~ <1-\frac{1}{2} Re~Tr~ U_{plaq}>~ =~ 1 - 
\frac{\beta}{4}+...  ~~~when~~~~\beta  \rightarrow 0
$$
and 
$$
P ~=~ \frac{3}{4\beta}+... ~~~~~~~when~~~~~~  \beta \rightarrow \infty.
$$

Let us now consider  two-plaquette action (\ref{linear}) which, as we shall 
see, has a different high temperature 
expansion. At high temperatures the average action is equal to:
$$ 
PP~=~  <1-\frac{1}{4} Re~Tr ~U_{plaq}\cdot 
Re~Tr ~U_{plaq}> ~=~1 - \frac{\beta}{16} +...
$$
and the low temperature expansion is the same as for the standard action
$$
PP = \frac{3}{4\beta}+...
$$
On Figure 1 one can see the behavior of average  
action in both models which we got by Monte Carlo simulations.
In the simulations we used the Heat Bath algorithm \cite{kenpen}. 
The new action
has indeed smaller values at high temperatures and the slope 
is four times smaller.

\begin{figure}
\centerline{\hbox{\psfig{figure=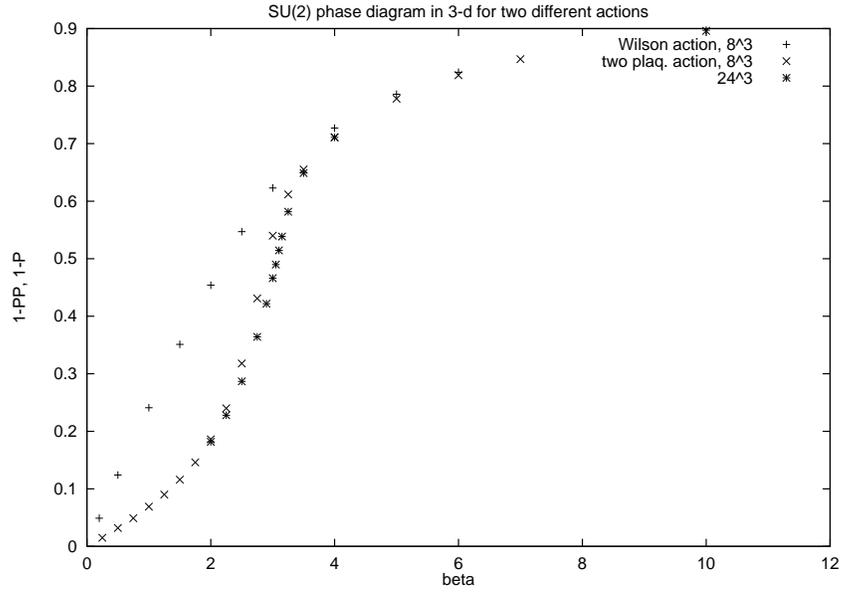,height=8cm,angle=-90}}}
\caption[fig1]{The behaviour of the average action in two different 
models. The new action has four times smaller values at high temperatures.}
\label{fig1}
\end{figure}
One can see also crossover from high to low temperatures. In the thermal 
cycles we have seen small hysteresis loop, but the measurement of susceptibility 
demonstrates that the maximum does not scale with the volume ( see Figure 2,3).
Thus we do not see any phase transition at $\beta \approx 3.5$.
\begin{figure}
\centerline{\hbox{\psfig{figure=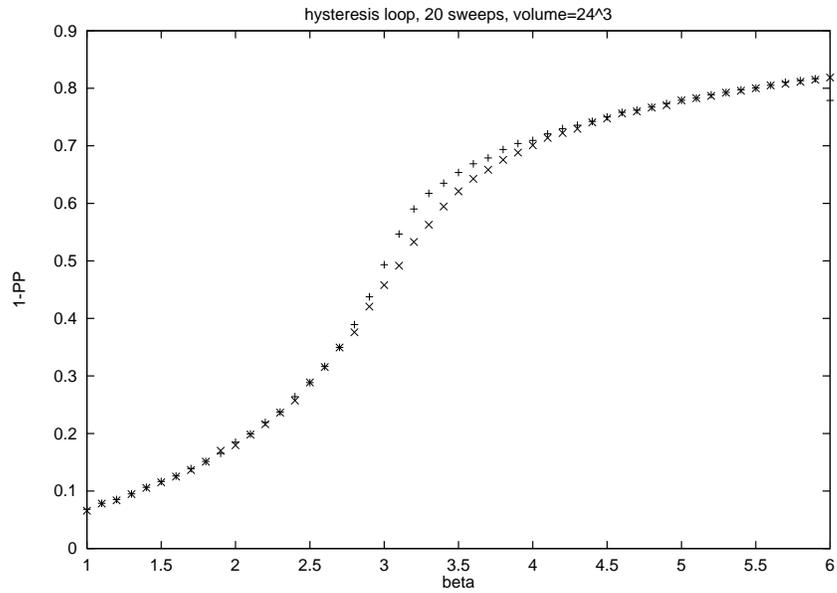,height=8cm,angle=-90}}}
\caption[fig2a]{The thermal cycle of the two-plaquette action.}
\label{fig2a}
\end{figure}

\begin{figure}
\centerline{\hbox{\psfig{figure=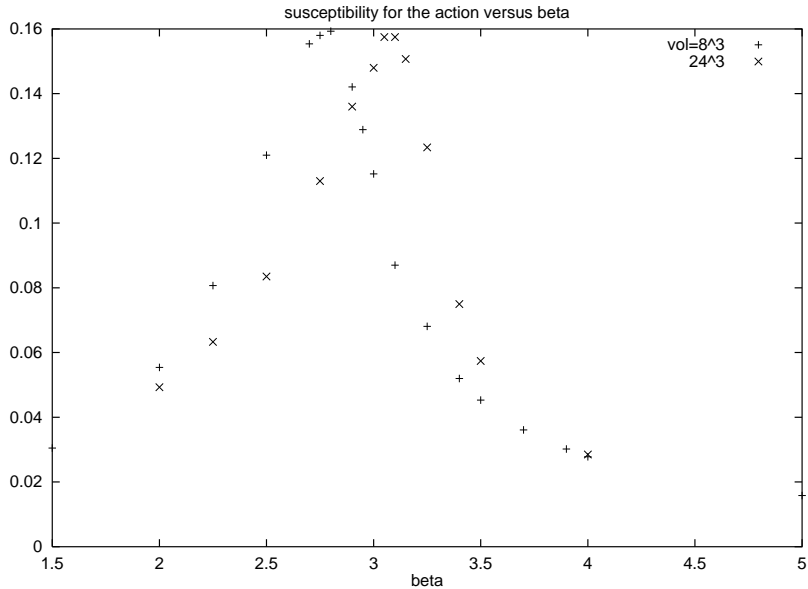,height=8cm,angle=-90}}}
\caption[fig2b]{Susceptibility versus $\beta$ for two volumes $8^3$,$24^3$. 
The maximum of susceptibility does not scale with the volume.}
\label{fig2b}
\end{figure}

\section{Loop averages and vacuum structure}

In the model with a one-plaquette action  (\ref{area}) the high temperature expansion 
of the loop averages  is equal to \cite{creutz}
$$
<W(C)>~ =~ W(I,J)~ =~ (\frac{\beta}{4})^{IJ},~~~~~~~~\beta \rightarrow 0,
$$
and if one defines $W(I,J) \approx exp(-\sigma A)$, where $A$ is the area $IJ$ 
of the loop $C$ then the string tension is equal to 
$$
\sigma ~\approx ~\frac{1}{a^2} ln (\frac{4}{\beta}).
$$

In the two-plaquette model  (\ref{linear}) the loop averages demonstrate peculiar 
behaviour. Indeed at high temperature the loop average
\be
<W(C)> = Z^{-} \cdot \int dU~  e^{-S}~~ \frac{1}{2}~ Tr_{C}~ U_{ij}
\ee
is equal to zero for $SU(2)$ gauge group, because there are no low order 
lattice diagrams (proportional, 
let us say, to area) which can contribute to the expectation value of $W(C)$
(see Figures 4 and 5). We have 
$absolute$ confinement in this model. It is similar to the limit $n \rightarrow \infty$ 
of the model with one-plaquette action 
$$
\sigma = \frac{1}{a^2} ln \frac{2n^2}{\beta} \rightarrow \infty
$$
and thus $<W(C)> =\approx exp(-\sigma A) \rightarrow 0$\footnote{The situation 
changes for large gauge groups where there
exist high temperature diagrams which contribute to one-plaquette average.
Specifically for the  $SU(3)$ group the 3D-box diagram on a top of plaquette 
gives nonzero contribution.}. 

The Monte-Carlo simulations of $SU(2)$ group demonstrate that this 
behaviour takes place until $\beta \
\approx 3.5 - 4$ as one can see on Figures 4 and 5. At these temperatures there 
appears proliferation of large lattice diagrams which cover almost 
the whole lattice and $<W(C)>$ gets nonzero value.

At low temperatures the nontrivial vacuum structure begins to play a 
central role. To see that let us consider the situation when two opposite 
plaquettes inside a 3D cube are frustrated. As it is easy to see from 
the action 
the energy of the new configuration will be the same. Indeed the energy
is minimal when $\frac{1}{4}Tr U_{plaq}\cdot Tr U^{paral}_{plaq} =1$, therefore we have 
two possibilities ether 
$$
\frac{1}{2}Tr~ U_{plaq} = \frac{1}{2}Tr~ U^{paral}_{plaq} =+1
$$ 
or 
$$
\frac{1}{2}Tr~ U_{plaq} = \frac{1}{2}Tr~ U^{paral}_{plaq} =-1.
$$ 
On the whole lattice one can build towers of frustrated plaquettes so 
that the new configuration will have the same energy as the trivial vacuum 
$\frac{1}{2}Tr~ U_{plaq} = +1$. Thus the vacuum has global degeneracy as it was in the 
corresponding $Z_2$ gauge spin system \cite{sav}. This actually means that 
we have many vacuua in which the gauge field $A_{\mu}$ is very far from 
perturbative  value.

\begin{figure}
\centerline{\hbox{\psfig{figure=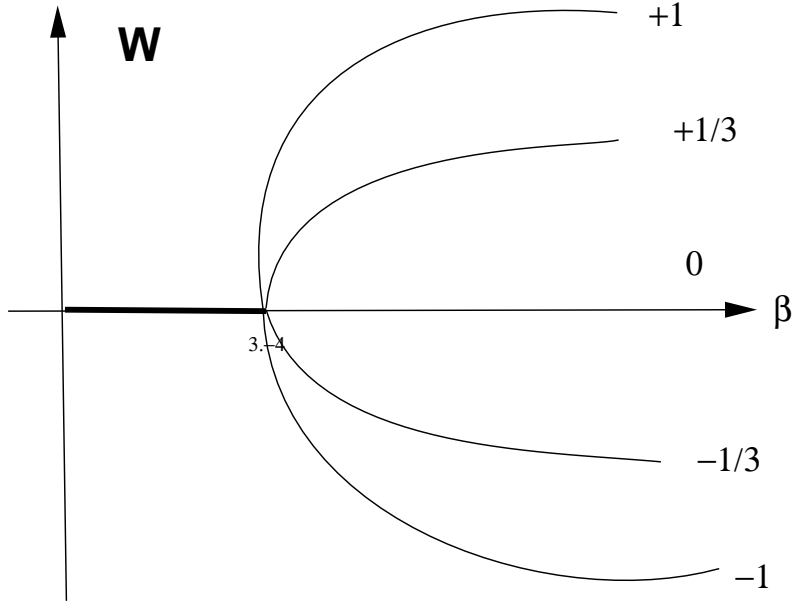,height=8cm,angle=00 }}}
\caption[fig3a]{Schematic multivacuum structure in the model with 
two-plaquette action. This figure shows the absolute confinement 
at high temperatures 
$\beta \leq 3.5$ and gauge nonequivalent vacuua at low temperatures.}
\label{fig3a}
\end{figure}

\begin{figure}
\centerline{\hbox{\psfig{figure=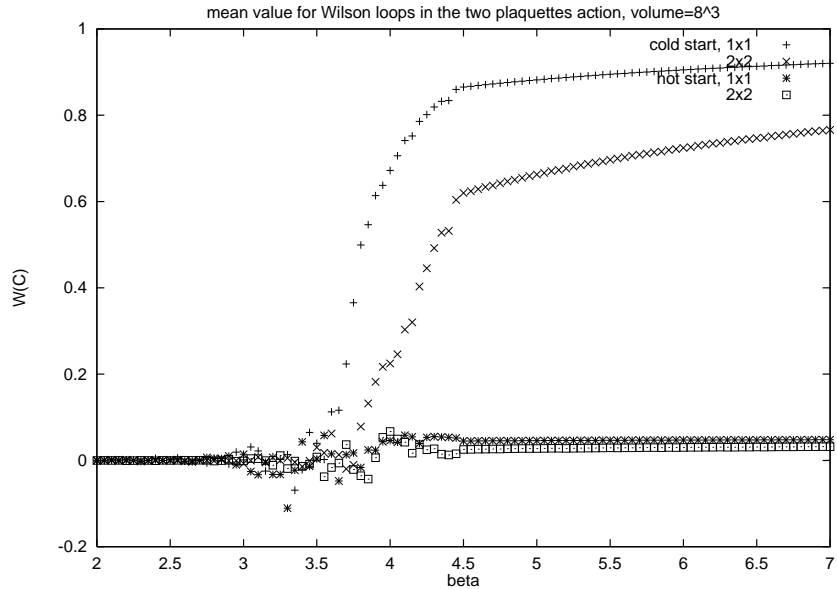,height=8cm,angle=-90}}}
\caption[fig3b]{ The behaviour of the mean value for the $1 \times 1$
and $2 \times 2$ loops for two-plaquette action. It represents the evolution of 
the averages starting from low temperature with $\frac{1}{2}Tr~U_{plaq} = +1$ 
to high 
temperature and then back to low one. As one can see the system fluctuates
at pseudocritical point and then falls into vacuum with 
$Tr~U_{plaq} \approx 0$ }
\label{fig3b}
\end{figure}

This complicated vacuum structure is reflected  on Figures 4 and 5 which show 
multivacuum structure at low temperature. The top curve corresponds to the 
vacuum in which all plaquettes are equal to plus one, $\frac{1}{2}Tr ~U_{plaq}=1$, and the 
bottom curve corresponds to the opposite case when all plaquettes are 
frustrated $\frac{1}{2}Tr~U_{plaq}=-1$.
It is also not difficult to construct explicitly vacuum configurations 
in which only part of the lattice plaquettes are frustrated.  
The vacuum with all plaquettes frustrated, $\frac{1}{2}Tr~U_{plaq}=-1$, 
corresponds to a gauge field which is far from the perturbative value 
$A_{\mu}=0$ in the 
whole lattice. 

Two natural questions arize:\\
i) is there a sensible continuum limit in all these vacuua in the limit $a \rightarrow 0,
\beta \rightarrow \infty$ ?\\
ii) and if there is a continuum limit, then what is difference between them ?
What type of parameter ("coupling constant") or field ("dilaton") may describe them?

\section{String tension in different vacuua}

In this paragraph we will study the behaviour of string 
tension in the limit $\beta \rightarrow \infty$. We choose this quantity because it can
characterize the system in the continuum limit if it exists.
As usually we define the string tension by the behaviour of W(L,L) for large L
as:
$$ 
W(L,L) \approx e^{-(\sigma L^2 + m L + Const)}
$$
To calculate string tension we shall follow the procedure: \\
i) calculate W(L,L) for different values of $L \leq 6$  at a
given $\beta $ (in our simulations the lattice volume is $16^3$ 
and the coupling constant is in the range $6 < \beta < 10$), \\
ii)then fit the quantity $-ln \vert W(L,L)\vert $ using second order polynomials and
extract $\sigma $ for each $\beta $, \\
iii)because string tension scales as $\sigma = \sigma_{phys} a^2$ and 
$\beta = 2/g^2 a$ we shall plot the ratio $\beta \sqrt{\sigma} = 
2 \sqrt{\sigma_{phys}}/ g^{2}$ versus $1/\beta$ (see Figure 6 ). \\

The last ratio converges to a finite limit when $\beta \rightarrow \infty$.
We expect that this limit can be reached within the finite size
corrections which are of order $O(\frac{1}{\beta})$.
On Figure 6 we also depicted  string tension in the vacuum 
$<\frac{1}{2}Tr~U_{plaq}> = +1$ calculated for the larger 
lattice $24^3$ in order to see the variation with volume. The convergence 
is better for small $\beta$.

Until now we have been calculating string tension 
in the vacuum $<\frac{1}{2}Tr~U_{plaq}> = +1$ when $\beta \rightarrow \infty$.
In the new vacuua we have frustrated plaquettes and the quantity 
$\frac{1}{2}Tr~U_{plaq}$~ is less than one. It remains constant during Monte Carlo
simulations for large $\beta \geq 6$ and we conclude that these vacuua are 
well separated \footnote{When we  
start from a given vacuum configuration  $<\frac{1}{2}Tr~U_{plaq}> \neq +1$
it never drifts into other vacuua when $\beta \geq 6$.}.

To explore vacuum structure we have to calculate $W(L,L)$
for these vacuua as well. 
A typical result is shown in  Table 1, where we choose $\beta = 10$,
the volume is equal to $16^3$ and L is smaller than six.
By "vacuum $-1/3$" we mean the vacuum which has the
mean value $ <\frac{1}{2} Tr~U_{plaq}> = -1/3$.

We observe that the absolute value $\vert W(L,L) \vert $ is the same   
for two vacuua +1 and  -1, as well for the vacuua +1/3 and -1/3.
The absolute value $\vert W(L,L) \vert $ for odd L in the vacuua $+1/3$ or $-1/3$
is  one third of its value in the vacuua $+1$ or $- 1$. 
Because of this  relation, 
 $\vert {W_{1}(L,L)} \vert = \frac {1}{3} \vert {W_{1/3}(L,L)}
\vert$
we have \footnote{By $W_{-1} (L,L)$ we mean the mean value of W(L,L) 
in the vacuum $Tr~U_{plaq}= -1$ for example.}
$$
  -ln\vert W_{1/3} (L,L) \vert = 
-ln \vert W_{1} (L,L) \vert  + ln(3)
   =\sigma L^2  + m L + ln(3) + Const
$$
and the string tension is the same for all these vacuua. 
Thus we can argue that string tension does indeed scale in all these vacuua 
and we can speculate that one can distinguish them by the scalar operator  
$\frac{1}{2}<Tr~U_{plaq}>$-gluon like condensate \cite{gs}. 

One of us (K.F.) wishes to thank the EU for partial financial support 
(TMR project FMRX-CT97-0122).

\begin{figure}
\centerline{\hbox{\psfig{figure=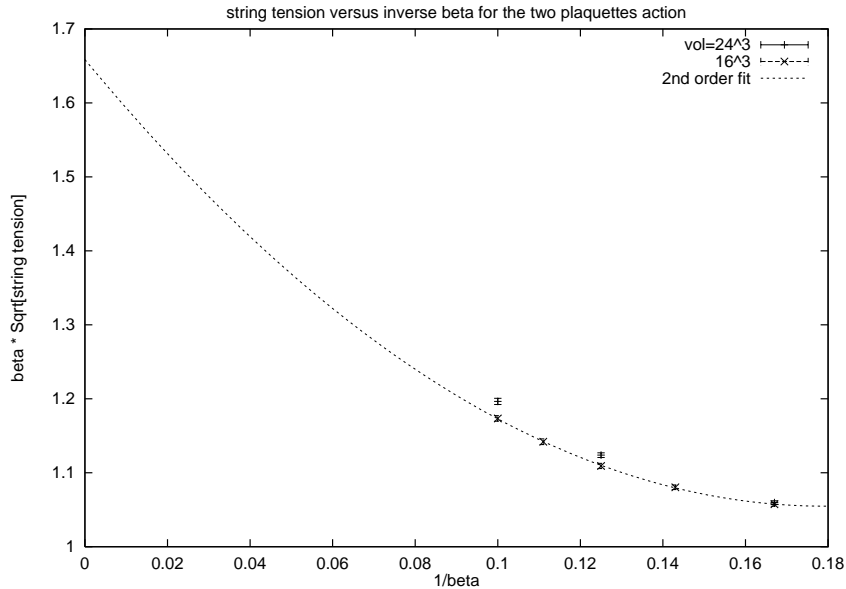,height=8cm,angle=-90}}}
\caption[fig4]{String tension versus inverse $\beta$ in the standard vacuum 
$Tr~U_{plaq} = +1$.}
\label{fig4}
\end{figure}

\begin{center}
\begin{tabular}{c}
\hline
\hline
TABLE 1\\
\hline
\hline
\\The values of $<W(L,L)>$ in different vacuua for $\beta =10$ and volume
 $16^3$ :\\
\end{tabular}
\end{center}
\begin{tabular}{||c|c|c|c|c|c|c||}
\hline
\hline
L&vacuum +1&vacuum -1& vacuum +1/3&vacuum- 1/3\\
\hline
\hline
 1&0.946281(4)&-0.946277(3)&0.315429(3)&-0.315427(4) \\
\hline
 2&0.83866(2)&0.83863(2)&0.83865(1)&0.83864(2) \\
\hline
 3&0.71894(4)&-0.71889(4)&0.23968(3)&-0.23967(4) \\
\hline
 4&0.60273(7)&0.60265(8)&0.60269(8)&0.60265(9) \\
\hline
 5&0.49675(13)&-0.49656(12)&0.16561(9)&-0.16567(10) \\
\hline
 6&0.40381(19)&0.40352(20)&0.40361(19)&0.40369(21)  \\
\hline
\hline
\end{tabular}

\vfill
\end{document}